# Community structure and modularity in networks of correlated brain activity

Adam J. Schwarz, Alessandro Gozzi, Angelo Bifone*

*Department of Neuroimaging, Psychiatry Centre of Excellence in Drug Discovery, GlaxoSmithKline, Via Fleming 4, 37135 Verona, Italy*
*Corresponding author (angelo.2.bifone@gsk.com)

**Abstract:** We present an approach to study functional segregation and integration in the living brain based on community structure decomposition determined by maximum modularity. We demonstrate this method with a network derived from functional imaging data with nodes defined by individual image pixels, and edges in terms of correlated signal changes. We found communities whose anatomical distributions correspond to biologically meaningful structures and include compelling functional subdivisions between anatomically equivalent brain regions.

The functional organization of the brains of higher vertebrates is based on the seemingly competing principles of segregation and integration [1,2]. Functional segregation has been observed at multiple scales, from local neuronal circuits to macroscopic, anatomically defined structures. At a microscopic level, for example, different neuronal groups respond preferentially to basic features of visual sensory inputs (e.g., orientation or color) [3]. At a larger scale, the cortex appears to be organized in spatially segregated areas associated with different functions, such as the processing of various sensory inputs [3,4]. The cortex itself is a functionally specialized structure, anatomically distinct from other parts of the brain that appeared at earlier stages of evolution. However, during complex activities of behavior and perception these specialized modules work in concert within functionally integrated, yet widely distributed, networks [5]. Indeed, complex behavior arises precisely due to the integrated action of different brain regions. This interplay between segregation and integration is central to understanding the emergence of higher brain function and ultimately of consciousness in the human brain [6].

Functional imaging methods, including functional Magnetic Resonance Imaging (fMRI), provide a powerful means to spatially and temporally resolve complex, dynamic patterns of brain activity in humans and laboratory animals [7]. Multivariate analysis can be applied to functional image time-series to examine correlated signal changes – across time, subjects or conditions – in different brain regions. Correlations between spatially remote events can be interpreted as *functional connectivity* [8], a concept that does not necessarily imply a direct correspondence with actual physical connections in the underlying neuronal substrate. Analyzing functional imaging data in this way, spatially distributed patterns of functional connectivity have been observed both in the resting state and in response to cognitive tasks or pharmacological challenge [9-11], consistent with the idea that brain function is highly integrated.

Functional connectivity patterns may also be represented in a very natural way as networks, with individual image voxels or anatomically defined structures representing the nodes and a measure of correlation between each pair determining the edges [12,13]. The coexistence of functional segregation and integration in brain activity [4,6] suggests that some degree of modularity might exist within the overall network. An issue that has received considerable recent attention is the identification of community structure [14] in complex networks [15] – the presence of clusters of nodes characterized by denser connections between their constituents than to other nodes outside the group. These concepts originated in the study of social relationships - hence the term "community" - but have recently begun to find application more broadly [14,16-19]. In this Letter, we apply a community structure algorithm based on maximization of modularity to investigate the structure of a real functional connectivity network derived from neuroimaging data. Specifically, we show that this approach can resolve meaningful functionally and anatomically segregated communities within widespread networks of correlated brain activity. Moreover, we discuss the interpretation of modularity in the context of functional connectivity, and argue that the community structure of a network representation of functional imaging data captures the interplay between segregation and integration in brain function.

We illustrate this approach with fMRI data acquired from the rat in which brain activity was stimulated by acute systemic administration of fluoxetine, an antidepressant drug that primarily affects the serotonergic system, one of the four major modulatory neurotransmitter systems and pervasive in the brain. Indeed, acute fluoxetine challenge results in a widespread response in the rat brain [10]. We have recently shown that the spatial profile of the response amplitude varies between subjects, enabling the calculation of functional connectivity maps based on inter-subject correlations [10]. Here we revisit this data set and represent it explicitly as a network, with the nodes defined as the individual pixels in the 3D image volume (0.128 mm$^3$ each) and the edges based on the Pearson correlation coefficient $r_{ij}$ of the response between each pair of nodes ($i,j$). This was converted into an equivalent $z$-statistic using Fisher's $r$-to-$z$ transformation:

$$z_{ij} = \left(\ln\left(1 + r_{ij}/1 - r_{ij}\right)\right) / \left(2\sqrt{1/(N-3)}\right), \tag{1}$$

where $N$ is the number of data points in the data vector associated with each node. The magnitudes of these normalized correlation values describe the strength of the correlation between each pair of nodes and can be used to construct a weighted adjacency matrix $W_{ij} = |z_{ij}|$, with edge weights reflecting the strength of functional coupling between nodes. For this data set, the resulting weighted complete network comprised 11492 nodes; for computational tractability, the connection strengths were thresholded to yield a binary adjacency matrix:

$$A_{ij} = \begin{cases} 1, & W_{ij} \geq z_0 \\ 0, & W_{ij} < z_0 \end{cases}. \tag{2}$$

We used a value of $z_0=3.5$, retaining the strongest 2% of the connections in the complete weighted network (on average, 115 edges per node). Structure within the full network, in the form of closely coupled communities of nodes reflecting sub-networks of brain regions, can be identified by finding a partition of the network that maximizes the *modularity* [14,18] $Q \in [-1,1]$, defined as

$$Q = \frac{1}{4m} \sum_{i,j} \left( A_{ij} - \frac{k_i k_j}{2m} \right) \delta_{ij}^{\text{group}} \quad , \tag{3}$$

where $m = \frac{1}{2}\sum_{i,j} A_{ij}$ is the total number of edges in the network, $k_i = \sum_j A_{ij}$ is the degree of node $i$ and $\delta_{ij}^{\text{group}}$ equals 1 if nodes $i$ and $j$ are in the same community and 0 otherwise. For a certain partition of the network, $Q$ measures the difference between the fraction of the edges connecting nodes within communities and the same fraction in the case of a randomly connected network with the same partition. The closer the value of $Q$ is to its theoretical maximum 1, the stronger the community structure, i.e. the more modular the network. Algorithms seeking a network partition that maximizes modularity yield an estimate of the number of communities into which the network should be optimally split, the composition of each community and an associated value of $Q$. In the case of functional connectivity networks, an optimal partition can reveal the system-level functional structure of the brain under the particular experimental conditions studied, as well as providing a measure of the emergent modularity. We explored community structure in the network activated by fluoxetine using the algorithm of Clauset *et al.* [16], which seeks the partitioning that maximizes the modularity using an agglomerative approach with only a linear dependence on network size.

Modularity in the fluoxetine network was maximized at $Q=0.38$ with a partition into three communities. The results are shown with the nodes in each community mapped back onto anatomical positions in three representative slices (Fig. 1). Encouragingly, the pixels in the two largest communities are symmetrically distributed between the left and right hemispheres, and their distributions correspond closely to known anatomical and functional subdivisions of the rat brain. It is important to emphasize that division was obtained from a network representation based on image pixels with no imposition of symmetry nor any prior anatomical constraints. The first community (Fig. 1(a); red) comprised nodes corresponding primarily to sub-cortical structures – in the striatum, thalamus and amygdala – but also included regions of the hippocampus and entorhinal, medial pre-frontal and cingulate cortices. The finding of pixels in the cingulate and prefrontal areas grouped with those in striatal and thalamic structures is a striking result – these cortical regions are anatomically identical to other regions of the cortex yet *functionally* distinct, being the main cortical target of input from the basal ganglia via extensive reciprocal connections with the thalamus [20]. This is consistent with our finding here of a community including these structures within the same functionally integrated unit. Similarly, pixels in the entorhinal cortex and in the hippocampus were assigned to the same community, reflecting the dense connections between these structures. Most other cortical nodes, including those located in the motor, somatosensory and visual cortices, were assigned to a second community (Fig. 1(b); blue), while a third community (Fig. 1(c); yellow) mainly comprised pixels near the brain edge, the ventricles and in white matter and the cerebellum.

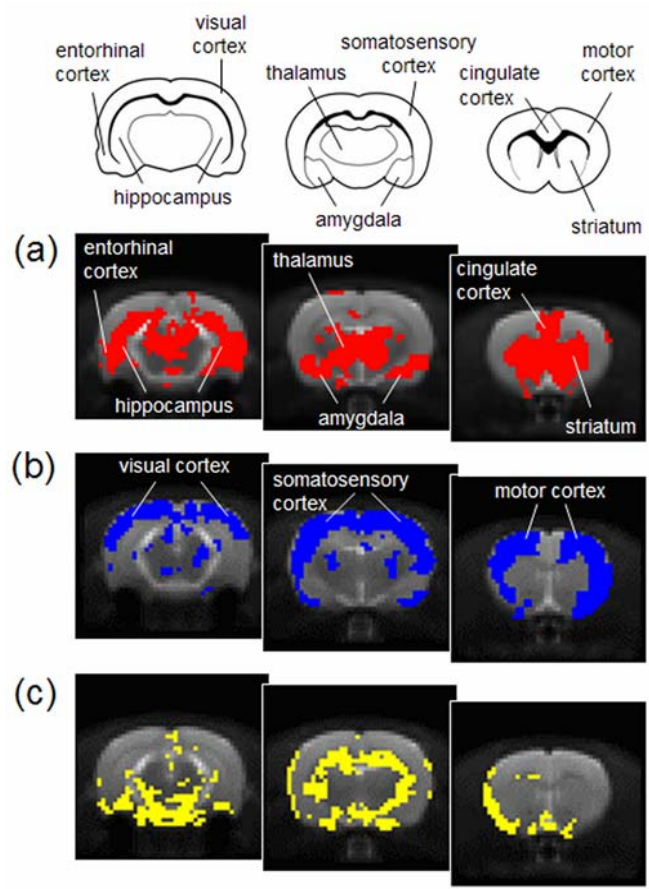

**Figure 1:** Anatomical representation of the three communities identified by modularity maximization (Q=0.38) in a fluoxetine functional imaging network (shown for three coronal slices). (a) A community comprising nodes corresponding primarily to sub-cortical structures in the striatum, thalamus and amygdala, along with the hippocampus and the entorhinal and pre-frontal and cingulate cortices (*N*=3912). (b) Most other cortical nodes were assigned to the blue community (*N*=4256). (c) An additional group comprises primarily nodes corresponding to pixels around the brain edge and ventricles (*N*=3324).

Since the algorithm assigns every connected node to a community, however, a community may merely represent a set of residual nodes, not especially tightly linked between themselves, rather than a group of strongly inter-correlated brain structures. To this end we explicitly examined the strength of coupling within each community compared with that between communities. A partition of a network $G$ assigns each node $i$ to a sub-network $V \subset G$. Comparing the within-group degree $k_i^{in}(V) = \sum_{j \in V} A_{ij}$ and between-group degree $k_i^{out}(V) = \sum_{j \notin V} A_{ij}$ for each node [19], we can consider the node-wise difference between $k^{in}$ and $k^{out}$, normalized to the number of nodes $N_V$ in the group, as a metric for each group $V$:

$$\Delta k_i(V) = \left(k_i^{in}(V) - k_i^{out}(V)\right) / N_V \qquad (4)$$

We expect that for meaningful communities, in which the connections within the sub-network are denser than those between sub-networks, the histograms will be substantially positive, equivalent to a right-shift in the $k^{in}$ histogram relative to $k^{out}$. In Fig. 2 we plot the distributions of $\Delta k$ for the three sub-networks identified above. In both the sub-cortical (red) and cortical (blue) sub-networks, the distributions are predominantly positive ($\Delta k > 0$ for 99% and 98% of the nodes, respectively, with the few nodes having negative $\Delta k$ corresponding to scattered pixels located at the edge of the maps shown in Fig. 1). Taken together with their symmetry and anatomical distributions, this indicates that both groups are meaningful communities of brain regions closely coupled in their response to the fluoxetine challenge. The third (yellow) sub-network appears qualitatively different, with its $\Delta k$ histogram less skewed and the distribution remaining closer to zero ($\Delta k > 0$ for 83% of the nodes), indicating that the nodes in this group are less tightly connected.

The approach presented here, of seeking community structure within a network representation of functional imaging data, is conceptually different to other methods usually applied to study segregation or integration in the brain. Most approaches have exploited differential responses to function-specific conditions inserted in the stimulation paradigm within a univariate framework, in which responses in different brain regions are analyzed independently [7]. Despite their great utility demonstrating functional specialization, these methods give no direct information regarding interactions between brain regions or how their activities are integrated. Moreover, these methods cannot be applied to cases such as the pharmacological challenge reported here where the stimulation paradigm cannot easily be manipulated. The present method is more akin to other multivariate methods that have been applied to seek structure within imaging data in a model-independent way [21-24]. Unlike these, however, our method explicitly takes into account the topology of the functional connections, with communities defined on the basis of link density and distribution. This concept is more readily interpretable in biological terms than measures such as the orthogonality [21,22] or statistical independence [23,24] of spatial modes that are optimized by other algorithms.

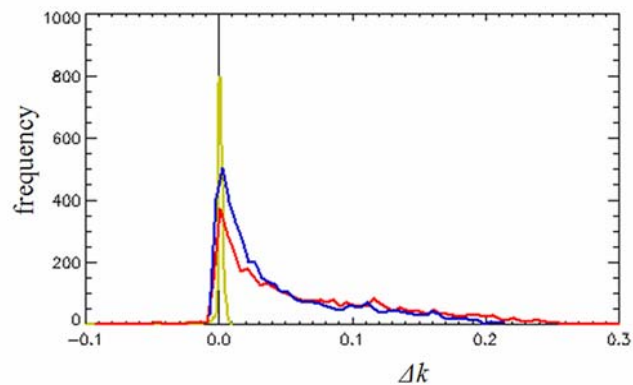

**Figure 2:** Histograms of $\Delta k$ for each of the three sub-networks identified in the pixellated network (colors corresponding to those in Fig. 1). The red and blue sub-networks have histograms skewed strongly positive, indicating denser within-group relative to between-group connections.

A key point in our approach is that the identification of communities within a functional imaging network contains information on both segregation *and* integration. A partition of the overall network into smaller sub-units suggests a degree of functional segregation in the response, whereas the set of brain regions – not necessarily contiguous – identified within each sub-network reflects their integrated action in response to the experimental stimulus. Importantly, the value of the modularity $Q$ provides a measure of the degree of functional segregation in the network and may provide the basis for an operational definition of this.

The results presented here were calculated on a binary version of the full functional connectivity network for reasons of computational tractability. Although the results of the partition are striking and biologically meaningful, the binary network itself represents an approximation with the full range of correlation strengths being lost. Networks based on a normalized correlation strength as defined herein represent a well-defined class of functional connectivity networks, applicable quite generally to functional imaging data (for example, to temporal correlations common in human functional connectivity analyses [12]). A number of recent studies indicate that more robust results can be obtained by retaining edge weight information [25,26]. However, functional connectivity networks retaining the full spatial resolution afforded by current neuroimaging techniques will typically comprise a large number of nodes ($\sim 10^4$-$10^5$). Determining the community structure in fully weighted networks of this size is computationally demanding, and development of efficient algorithms will allow full exploitation of the rich information contained therein. Also, a more quantitative interpretation of the modularity parameter $Q$ awaits a rigorous definition of an appropriate null model for the specific class of networks considered.

In conclusion, we have applied a community structure approach based on maximization of modularity to a real biological network derived from correlated signal changes in the living brain. The anatomical and functional specificity of the results shown here provide compelling evidence that community structure algorithms can generate biologically meaningful partitions of networks based on functional imaging data. Moreover, the resulting $Q$ value represents a measure of the degree of residual modularity in the network, thus providing important information regarding the interplay between integration and segregation in brain function. The network we analyzed is one of the brain activated following a systemic pharmacological challenge, but this approach is straightforwardly applicable to functional connectivity networks arising from other pharmacological or cognitive stimuli, to human or pre-clinical species, at single-subject or group level – as long as a measure of functional connection can be defined.